\documentclass[prb,a4paper,showpacs,superscriptaddress,twocolumn]{revtex4}
\usepackage{amsmath}
\usepackage{amssymb}
\usepackage{amsfonts}
\usepackage{graphicx}
\usepackage{bm}

\DeclareMathOperator{\sgn}{sgn} 
\DeclareMathOperator{\Real}{Re}

\begin{document}

\title{Surface states in a 3D topological insulator: The role of hexagonal warping and curvature}

\author{E.V. Repin}
\affiliation{Moscow Institute of Physics and Technology, 141700 Moscow, Russia}
\author{V.S. Stolyarov}
\affiliation{Moscow Institute of Physics and Technology, 141700 Moscow, Russia}
\affiliation{Sorbonne Universit\'{e}s, UPMC Univ Paris 06, UMR 7588, Institut des Nanosciences de  Paris, F-75005, Paris, France}
\affiliation{CNRS, UMR 7588, Institut des Nanosciences de Paris, F-75005, Paris, France}
\affiliation{Institute of Solid State Physics RAS, 142432, Chernogolovka, Russia}

\author{T. Cren}
\affiliation{Sorbonne Universit\'{e}s, UPMC Univ Paris 06, UMR 7588, Institut des Nanosciences de Paris, F-75005, Paris, France}
\affiliation{CNRS, UMR 7588, Institut des Nanosciences de Paris, F-75005, Paris, France}

\author{C. Brun}
\affiliation{Sorbonne Universit\'{e}s, UPMC Univ Paris 06, UMR 7588, Institut des Nanosciences de Paris, F-75005, Paris, France}
\affiliation{CNRS, UMR 7588, Institut des Nanosciences de Paris, F-75005, Paris, France}

\author{S.I.~Bozhko}
\affiliation{Institute of Solid State Physics RAS, 142432, Chernogolovka, Russia}

\author{L.V. Yashina}
\affiliation{Department of Chemistry, Moscow State University, Leninskie Gory 1/3, 119991, Moscow, Russia}

\author{D. Roditchev}
\affiliation{Sorbonne Universit\'{e}s, UPMC Univ Paris 06, UMR 7588, Institut des Nanosciences de Paris, F-75005, Paris, France}
\affiliation{CNRS, UMR 7588, Institut des Nanosciences de Paris, F-75005, Paris, France}
\affiliation{LPEM, ESPCI ParisTech-UPMC, CNRS-UMR 8213, 10 rue Vauquelin, 75005 Paris, France}
\author{I.S. Burmistrov}
\affiliation{L.D. Landau Institute for Theoretical Physics RAS, Kosygina street 2, 119334 Moscow, Russia}
\affiliation{Moscow Institute of Physics and Technology, 141700 Moscow, Russia}

\begin{abstract}
We explore a combined effect of hexagonal warping and of finite effective mass on both the tunneling density of electronic states (TDOS) and structure of Landau levels (LLs) of 3D topological insulators. We find the increasing warping to transform the square-root van Hove singularity into a logarithmic one. For moderate warping an additional logarithmic singularity and a jump in the TDOS appear. This phenomenon is experimentally verified by direct measurements of the local TDOS in Bi$_2$Te$_3$. By combining the perturbation theory and the WKB approximation we calculate the LLs in the presence of hexagonal warping. We predict that due to the degeneracy removal the evolution of LLs in the magnetic field is drastically modified.
\end{abstract}


\maketitle


\section{Introduction}

Theoretical and experimental study of three dimensional (3D) topological insulators is in the focus of modern research in condensed matter physics.~\cite{HK,QZ,Ando2013} Apart from fundamental interest to the novel quantum state of matter topological insulators attract a lot of attention provoked by their possible applications in spintronics due to spin-current locking of surface states. Many exciting features of electron states on the surface of a 3D topological insulator were found within the simplest two dimensional (2D) hamiltonian linear in momentum and spin operators which is allowed by the time-reversal and crystal symmetries.~\cite{HK,QZ,Ando2013}

Recently it was realized that without violation of the symmetry this simplest hamiltonian can be extended to higher order terms in momentum describing finite mass and hexagonal warping of surface states.~\cite{F,LQ} Indeed, the hexagonal warping of their Fermi surface has been found experimentally by angle resolved photoemission spectroscopy (ARPES) in such topological insulators as Bi$_2$Te$_3$, \cite{Chen2009,ZA1}, Bi$_2$Se$_3$, \cite{Kuroda2010} and Pb(Bi,Sb)$_2$Te$_4$. \cite{NS} Theoretically, the hexagonal warping of the surface states can induce spin-density wave instability, \cite{F} affects the dc and optical conductivities, \cite{Wang2011,Xiao2013} is responsible for localization of the Cherenkov sound in certain directions, \cite{Smirnov2013} and can stabilize the $\nu=1/3$ fractional quantum Hall state. \cite{Fu2013} In addition to the hexagonal warping the spin and angle resolved photoemission spectroscopy revealed the presence of finite curvature of the spectrum of surface states in Bi$_2$Te$_3$, Bi$_2$Se$_3$, Pb(Bi,Sb)$_2$Te$_4$ and TlBiSe$_2$. \cite{NS}

Alternative experimental way to access the spectrum of surface states in 3D topological insulators is the scanning tunneling microscopy. Recently scanning tunneling microscopy was employed for Bi$_2$Te$_3$, \cite{ZA1,Urazhdin,ZA2,SO} Bi$_2$Se$_3$, \cite{Urazhdin,Hanaguri2010,tZhang2013,YSFu2014,Zaitsev2014} and Sb$_2$Te$_3$ in a perpendicular magnetic field. \cite{YY} The spectrum of surface states extracted from ARPES data is correlated with the tunneling conductance measured by scanning tunneling microscopy. \cite{ZA1} However, bulk states contribute also to
the tunneling conductance thus hiding a part due to the surface states. In order to unravel the surface contribution it is crucial to know the tunneling density of surface states (TDOSS) in detail. Within the spectrum linear in momentum the TDOSS with and without magnetic field was studied theoretically in Refs. [\onlinecite{Saha2011,Schwab2012,Vazifeh2012}]. In spite of clear experimental relevance, we are not aware of theoretical studies of the TDOSS in the presence of non-zero curvature and hexagonal warping.

In this paper we calculate the tunneling density of states on the surface of 3D topological insulator in the presence of hexagonal warping and finite mass $m$. We demonstrate that hexagonal warping leads to logarithmic van Hove singularity instead of the square-root one which exists in the case of a finite mass due to the end point of the spectrum. For moderate values of the hexagonal warping we discover additional logarithmic singularity and  a jump in the TDOSS. This prediction is quantitatively supported by scanning tunneling microscopy measurements of the local density of states in Bi$_2$Te$_3$. In the presence of perpendicular magnetic field we analyze structure of Landau levels within the perturbation theory and in the WKB approximation. As well-known, \cite{BR} in the absence of hexagonal warping there are crossings of Landau levels at some magnetic fields due to a finite mass. We find that the hexagonal warping removes these degeneracies and strongly affects the slope of Landau levels with respect to magnetic field. 

The paper is organized as follows. In Sec. \ref{sec1} we introduce the model hamiltonian and calculate
the tunneling density of states on the surface of 3D topological insulator in the presence of hexagonal warping and finite mass.  In Sec. \ref{sec2} we analyze the effect of hexagonal warping on Landau levels within the perturbation theory. In Sec. \ref{sec3} we investigate structure of Landau levels in the presence of hexagonal warping in the WKB approximation. In Sec. \ref{sec4} we report experimental results for the local density of states. We conclude the paper with discussion of how our theoretical results can be further tested experimentally (Sec. \ref{sec5}).

\section{Tunneling density of surface states at zero magnetic field \label{sec1}}

We start from the model hamiltonian of electron states on the surface of 3D topological insulator in zero magnetic field which is the following $2\times 2$ matrix: \cite{F,LQ}
\begin{equation}
\mathcal{H} = v \bigr (k_{x}\sigma_{y}-k_{y}\sigma_{x}\bigl )+\frac{k_x^2+k_y^2}{2m}+\frac{\lambda}{2}\bigr ( k_{+}^3+k_{-}^3\bigl )\sigma_{z} .
\label{eq:ham1}
\end{equation}
Here $\bm{k}=\{k_x, k_y\}$ denotes in-plane quasiparticle momentum, $k_{\pm}=k_x\pm i k_y$, and
$\sigma_{x,y,z}$ are the Pauli matrices. We note that due to spin-orbit coupling in the bulk the Pauli matrices $\sigma_{x,y,z}$ do not necessary correspond to operators of the electron spin. \cite{Silvestrov2012,fZhang2012} The first term in the right hand side of Eq. \eqref{eq:ham1} describes the conical (Dirac-type) spectrum with  velocity $v$. The second term in Eq. \eqref{eq:ham1} takes into account a finite curvature of the surface state spectrum. An effective mass $m$ can be positive (e.g., for Bi$_2$Se$_3$) or negative (as in the case of Bi$_2$Te$_3$). \cite{NS} In what follows, having in mind the case of Bi$_2$Te$_3$, we consider the situation of $m<0$. The results for the opposite case, $m>0$, can be easily obtained by inversion of the energy and momentum. The last term of Eq. \eqref{eq:ham1} describes the effect of the hexagonal warping whose strength is characterized by the parameter
 $\lambda$. In the absence of the hexagonal warping, $\lambda=0$, the hamiltonian \eqref{eq:ham1} is just the Bychkov-Rashba hamiltonian for 2D electrons with spin-orbit splitting. \cite{BR}  One can add to the hamiltonian \eqref{eq:ham1} the term of the third order in momentum describing the $\bm{k}^2$ contribution to the velocity $v$. \cite{F} Moreover, extension of the hamiltonian \eqref{eq:ham1} to the fifth order in $k$ Dresselhaus spin-orbit terms was proposed to explain deviation of the electron spin from the direction perpendicular to the momentum. \cite{Basak2011} However, recent results of spin and angle resolved photoemission spectroscopy \cite{NS} do not demonstrate significant deviation of the surface state spectrum from one corresponding to Eq. \eqref{eq:ham1}. Therefore, we confine our considerations to the hamiltonian \eqref{eq:ham1}.

\begin{figure}[t]
\centerline{(a) \includegraphics[width=0.2\textwidth]{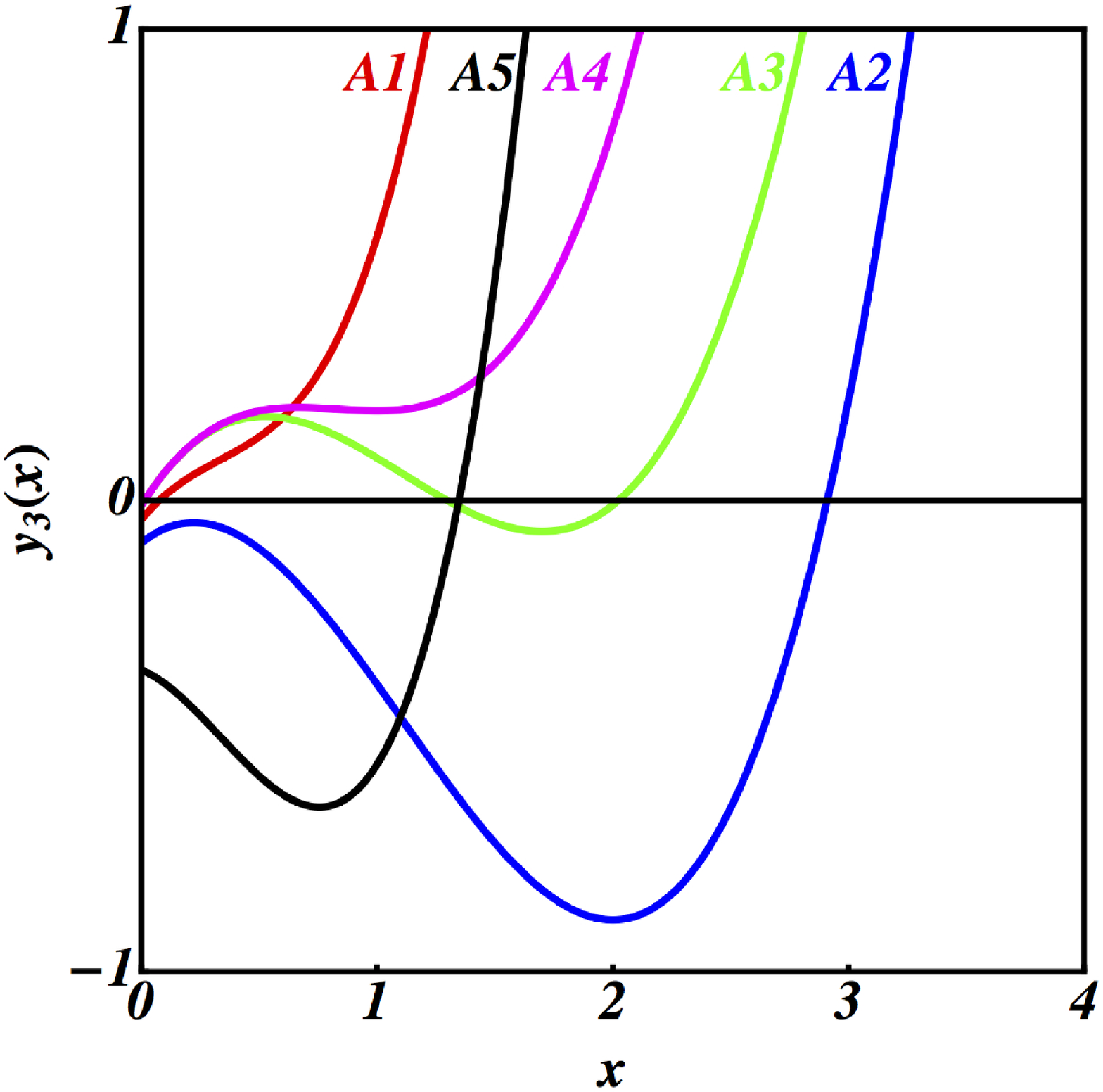} \quad (b) \includegraphics[width=0.2\textwidth]{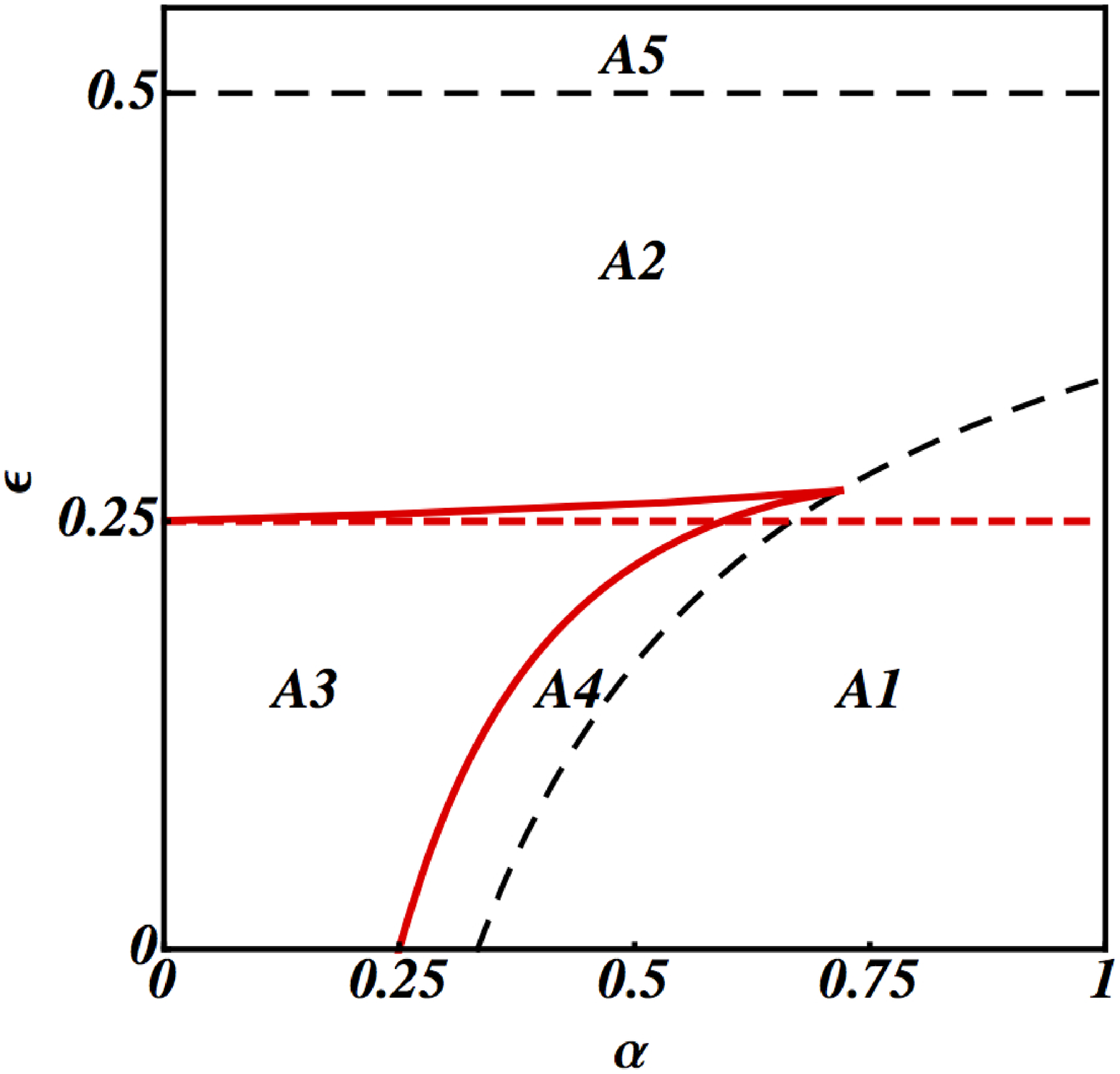}}
\caption{(Color online) (a) The five different types of possible behavior for the cubic polynomial $y_3(x)$. (b) The five corresponding regions in the $\{\epsilon,\alpha\}$ plane.}
\label{Fig:Fig1}
\end{figure}

The spectrum of the hamiltonian \eqref{eq:ham1} has the following form \cite{F,LQ}
\begin{equation}
E_\pm(k,\theta)=\frac{k^2}{2m}\pm \sqrt{v^2k^2+\lambda^2 k^6 \cos^2 3 \theta} ,
\label{eq:spectrum1}
\end{equation}
where $\theta$ parameterizes the momentum, $k_x=k\cos\theta$, $k_y=k \sin\theta$. The TDOSS can be written as
\begin{equation}
g(E)=\sum_{s=\pm}^{}\int\limits_{0}^{\infty} \frac{k dk}{(2\pi)^2}\int\limits_{0}^{2\pi}d\theta\, \delta \bigl (E - E_s(k,\theta)\bigr ) .
\label{eq:dos}
\end{equation}
It is convenient to introduce the energy parameters $E_0=\sqrt{v^3/\lambda}$ and $\Delta=2|m|v^2$ to characterize the hexagonal warping and curvature, respectively.
Then the dimensionless parameter $\alpha=({\Delta}/{E_0})^4$ measures the strength of hexagonal warping in comparison with the curvature. We remind that in the absence of warping, $\alpha=\lambda=0$, the density of states reads
\begin{equation}
g_{\lambda=0}(E)=\frac{\Delta}{2\pi v^2}
\begin{cases}
1, \, & E < 0 , \\
(1-4E/\Delta)^{-1/2}, \, & 0\leqslant E < \Delta/4 ,\\
0, \, & \Delta/4 < E .
\end{cases}
\end{equation}
It has the square-root van Hove singularity at $E=\Delta/4$ which is the end point of the spectrum.
For non-zero hexagonal warping, $\alpha>0$,  the TDOSS is given as
\begin{equation}
g(E)=\frac{\Delta}{2\pi v^2}F(E/\Delta,\alpha)  ,
\end{equation}
where the function
\begin{align}
F(\epsilon,\alpha)= & \frac{1}{\pi}  \int\limits_{0}^{\infty} dx\, |\epsilon+x| \Real\frac{1}{\sqrt{(\epsilon+x)^2-x}}\notag \\
& \times  \Real\frac{1}{\sqrt{\alpha x^3+x-(\epsilon+x)^2}} .
\label{eq:DefF}
\end{align}
Limits of integration over $x$ in Eq. \eqref{eq:DefF} are determined, in fact, by the regions where radicands are positive. Depending on values of $\epsilon$ and $\alpha$ the cubic polynomial $y_3(x) = \alpha x^3+x-(\epsilon+x)^2$ can have one (see curves A1, A2, A4, A5 on Fig. \ref{Fig:Fig1}a) or three (see curve A3 on Fig. \ref{Fig:Fig1}a) real roots. The regions of corresponding behavior in the $\{\epsilon,\alpha\}$ plane are shown in  Fig. \ref{Fig:Fig1}b.
There is the region A5 above the line $\epsilon=1/2$. The region A1 is situated below the curve $\alpha_1(\epsilon)=1/[3(1-2\epsilon)]$. The region A3 is clamped between the curves parameterized as $\alpha=\alpha_-(\epsilon)$ and $\alpha=\alpha_+(\epsilon)$ where
\begin{gather}
\alpha_\pm(\epsilon) = \frac{2(\epsilon+z_\pm(\epsilon))-1}{3z_\pm^2(\epsilon)}, \notag \\
 z_\pm(\epsilon)=1-2\epsilon\pm \sqrt{(1-2\epsilon)^2-3\epsilon^2} .
\end{gather}
The curves $\alpha=\alpha_\pm(\epsilon)$ are merged and end at the point $\epsilon_c={1}/({2+\sqrt{3}}) \approx 0.27$ and $\alpha_c=(3+2\sqrt{3})/{9}\approx 0.71$. The region A2 is below the region A5 but above the curve parameterized as $\alpha = \max\{\alpha_-(\epsilon),\alpha_1(\epsilon)\}$. The region A4 is clamped between the curves $\alpha=\alpha_+(\epsilon)$ and $\alpha=\alpha_1(\epsilon)$.

Let us denote the roots of the cubic polynomial $y_3(x)$ in order of increase as $c_1, c_2, c_3$, if there exist three real roots, and $c_i$, where $i=$ 1 or 3, in the case of a single real root only. We note that $z_-$ ($z_+$) coincides with $c_1$ and $c_2$ ($c_2$ and $c_3$) at the point where they merge. The roots of the quadratic polynomial $y_2(x)=(\epsilon+x)^2-x$ are given as $x_{1,2} = (1-2\epsilon \mp\sqrt{1-4\epsilon})/2$. It is convenient to introduce the following functions
\begin{gather}
F_1 = \int\limits_{c_1}^{x_1} dx \, \mathcal{F}(x,\epsilon,\alpha), \quad
F_2 = \int\limits_{x_2}^{c_2} dx \, \mathcal{F}(x,\epsilon,\alpha),\notag \\
F_3 =\int\limits_{c_3}^{\infty} dx \, \mathcal{F}(x,\epsilon,\alpha), \quad
F_4 = \int\limits_{c_1}^{c_2} dx \, \mathcal{F}(x,\epsilon,\alpha), \notag \\
F_5 = \int\limits_{x_2}^{\infty} dx \, \mathcal{F}(x,\epsilon,\alpha) ,
\end{gather}
where
\begin{equation}
\mathcal{F}(x,\epsilon,\alpha) = \frac{1}{\pi} \frac{|\epsilon+x|}{\sqrt{(\epsilon+x)^2-x}\sqrt{\alpha x^3+x-(\epsilon+x)^2}} .
\end{equation}
Then for each region in Fig. \ref{Fig:Fig1}b the function $F(\epsilon,\alpha)$ can be represented as a linear combination of functions $F_i$, $i=1,\dots 5$ with coefficients equal to $0$ or $1$ (see Table \ref{TabF}).

The TDOSS has singular behavior on the line $\epsilon=1/4$ and on the curves $\alpha=\alpha_\pm(\epsilon)$.
The logarithmic divergence at $\epsilon=1/4$ for any $\alpha>0$ is successor of the square-root singularity at the same energy existing in the case $\alpha=0$. Formally, it is due to consolidation of two real roots $x_{1,2}$ of the quadratic polynomial $y_2(x)$. The asymptotic of $F(\epsilon,\alpha)$ near this logarithmic singularity is as follows
\begin{equation}
F(\epsilon,\alpha)
\approx \frac{4}{\pi \sqrt{\alpha}}\ln \frac{1}{|\epsilon-1/4|} , \qquad \left |\epsilon- {1}/{4}\right | \ll 1 .
\label{eq:a1}
\end{equation}

There is the other logarithmic divergence of the density of states at the curve $\alpha=\alpha_+(\epsilon)$. Within the logarithmic accuracy the asymptotic behavior of the function $F$ near $\alpha=\alpha_+(\epsilon)$ can be found as
\begin{equation}
F(\epsilon,\alpha) \approx
\frac{\Delta_+}{\pi} \ln\frac{1}{|\alpha-\alpha_+(\epsilon)|} ,\qquad  \left |\alpha-\alpha_+(\epsilon)\right | \ll 1 ,
\label{eq:a2}
\end{equation}
where
\begin{equation}
\Delta_+ = \frac{|\epsilon+z_+(\epsilon)|}{[(\epsilon+z_+(\epsilon))^2-z_+(\epsilon)]^{1/2}[1-3(1-2\epsilon)\alpha_+(\epsilon)]^{1/4}} .
\label{eq:Delta:+}
\end{equation}
At the border between regions A2 and A3 there is a jump of the density of states due to appearance of infinitely small range of integration between the first two roots $c_1$ and $c_2$ of the cubic polynomial $y_3(x)$. We find for the jump of the function $F(\epsilon,\alpha)$ at $\alpha=\alpha_-(\epsilon)$
\begin{gather}
F(\epsilon,\alpha_--0) - F(\epsilon,\alpha_-+0) =\lim\limits_{c_1\to c_2} \int_{c_1}^{c_2}\frac{dx}{\pi \sqrt{\alpha}}  \notag \\
\times \frac{|\epsilon+x| [(\epsilon+x)^2-x]^{-1/2}}{[(c_3-x)(x-c_1)(c_2-x)]^{1/2}} = \Delta_- .
\label{eq:a3}
\end{gather}
Here $\Delta_-$ is given by Eq. \eqref{eq:Delta:+} after the substitution of $z_-$ and $\alpha_-$ for $z_+$ and $\alpha_+$, respectively.

\begin{table}[t]
\caption{The expressions for the function $F(\epsilon,\alpha)$ in different regions of the $\{\epsilon,\alpha\}$ plane. (see text)}
\begin{tabular}{c|clcl}
& \hspace{0.2cm} & $\epsilon< 1/4$ &\hspace{0.2cm}  & $\epsilon \geqslant 1/4$ \\
\hline
A1 & & $F= F_1+F_5$ & &  $F= F_3+F_4$ \\
A2 & & & & $F=F_3$\\
A3 & & $F= F_1+F_2+F_3$ & & $F=F_3+F_4$ \\
A4 & & $F= F_1+F_5$ & & $F= F_3+F_4$ \\
A5 & & & & $F=F_3$
\end{tabular}
\label{TabF}
\end{table}

\begin{figure}[t]
\centerline{\includegraphics[width=0.35\textwidth]{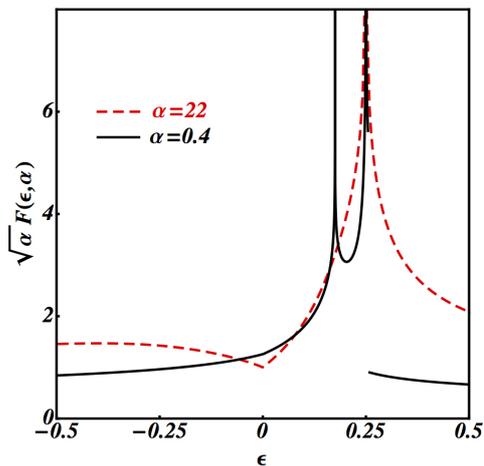}}
\caption{(Color online) The normalized TDOSS versus dimensionless energy $\epsilon=E/\Delta$ for $\alpha=0.4$ (solid black curve) and $\alpha=22$ (dashed red curve).}
\label{Fig:Fig2}
\end{figure}

Therefore, for $\alpha>0$  the square-root divergence of the density of states at $E=\Delta/4$ is split into the logarithmic divergence and the jump. The latter exists for $\alpha<\alpha_c$ only. The second logarithmic divergence appears from $\epsilon=-\infty$ with increase of $\alpha$ from zero value. Such nontrivial behavior of the TDOSS (the function $F(\epsilon,\alpha)$) is illustrated in Fig. \ref{Fig:Fig2}.

\begin{figure}[t]
\centerline{\includegraphics[width=8cm]{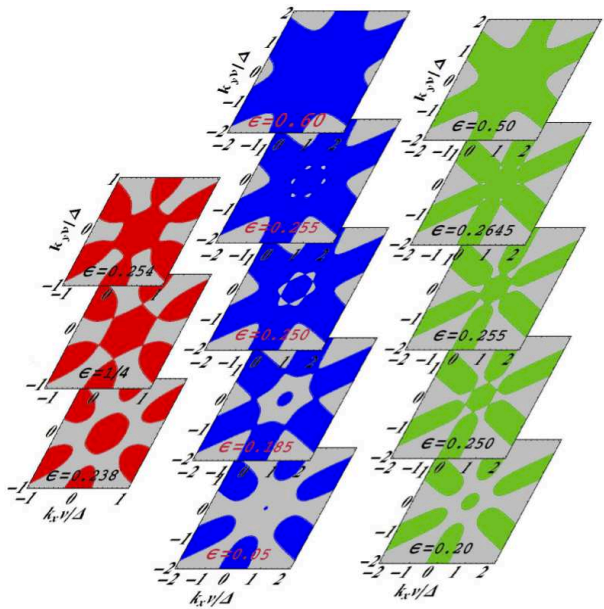}}
\caption{(Color online) The constant energy cuts of the energy spectrum \eqref{eq:spectrum1} for
(a) $\alpha=1$, (b) $\alpha=0.4$ and (c) $\alpha=0.68$.}
\label{Fig:Fig3}
\end{figure}

As usual, the van Hove singularities in the density of states discussed above can be explained by a complicated, not linearly connected shape of a Fermi surface for the spectrum, Eq. \eqref{eq:spectrum1}. The Fermi surface is illustrated graphically in Fig. \ref{Fig:Fig3}. Depending on the values of $\alpha$ there are three different cases of possible evolution of the Fermi surface with increase of the chemical potential (energy).  In the case $\alpha>\alpha_c$ there is one logarithmic divergence of the density of states at $E=\Delta/4$. It is due to touching of  the central snowflakelike part enclosing the $\bar{\Gamma}$ point and the six outermost disconnected parts (see Fig. \ref{Fig:Fig3}a, panel with $\epsilon=0.250$). For $\alpha<\alpha_0$ where $\alpha_0 = \alpha_+(1/4) = 16/27 \approx 0.59$, two logarithmic singularities
exist in the density of states. The first one at $E= \epsilon_+ \Delta$ ($\epsilon_+$ is determined as the solution of the following equation: $\alpha=\alpha_+(\epsilon_+)$) is related to touching of the six outermost disconnected parts with each other (see Fig. \ref{Fig:Fig3}b, panel $\epsilon=0.185$). The second singularity situated at $E=\Delta/4$ is due to touching of the central snowflakelike part and the part formed after consolidation of six initially disconnected pieces (see Fig. \ref{Fig:Fig3}b, panel $\epsilon=0.250$). The jump in the density of states at $E= \epsilon_- \Delta$ ($\epsilon_-$ is determined as the solution of the following equation: $\alpha=\alpha_-(\epsilon_-)$) is related to disappearance of six empty spots (see Fig. \ref{Fig:Fig3}b, panel $\epsilon=0.255$).
In the intermediate range, $\alpha_0 < \alpha \leqslant \alpha_c$, there are two logarithmic singularities of the density of states. The first one at $E=\Delta/4$ is  due to touching of the central snowflakelike part and the six outermost disconnected parts (see Fig. \ref{Fig:Fig3}c, panel $\epsilon=0.250$). The second singularity at $E=\epsilon_+\Delta $ is related to appearance of six empty spots (see Fig. \ref{Fig:Fig3}c, panel $\epsilon=0.255$). The jump in the density of state is due to collapse of these empty spots (see Fig. \ref{Fig:Fig3}c, panel $\epsilon=0.2645$).

\section{Landau levels within perturbation theory \label{sec2}}

Now we consider the effect of magnetic field $H$ perpendicular to the surface of a 3D topological insulator on the spectrum of surface states. In general, one needs to start from the hamiltonian describing bulk states in the presence of magnetic field and to derive from it the effective 2D hamiltonian for the surface states. It was shown \cite{YH} that such approach leads to the results which are similar to the results that can be obtained from the zero-field hamiltonian for the surface states after the Peierls substitution. Therefore, to describe the surface states in perpendicular magnetic field we substitute the momentum $\bm{k}$  in the hamiltonian \eqref{eq:ham1} by  $\bm{k} - e \bm{A}$. Here  $\bm{A}$ denotes the vector potential for the perpendicular magnetic field, $\bm{H} = \nabla\times \bm{A}$, and $e$ stands for the electron charge. In addition the Zeeman term $g_L \mu_B H \sigma_z/2$ ($g_L$ and $\mu_B$ are the $g$-factor and Bohr magneton, respectively) has to be added to the hamiltonian \eqref{eq:ham1}. Here we assume for simplicity the (111) surface such that $\bm{\sigma}/2$ coincides with the electron spin operator. \cite{Silvestrov2012,fZhang2012} Thus we consider the following hamiltonian:
\begin{gather}
\mathcal{H} = \frac{(\bm{k}-e\bm{A})^2}{2m}+v \bigl [(\bm{k}-e\bm{A}),\bm{\sigma}\bigr ]_{z}+\frac{\lambda}{2} \sum_{s=\pm} (k_{s}-eA_s)^3\sigma_{z}\notag \\
 + \frac{1}{2} g_L \mu_B H \sigma_z ,
\label{eq:ham1:mag}
\end{gather}
where $A_\pm=A_x \pm iA_y$. For the case $\lambda=0$ the hamiltonian \eqref{eq:ham1} describes 2D electrons with Rashba-type spin-orbit splitting in the presence of magnetic field. \cite{BR} Then the spectrum (Landau levels) are known to be as follows: \cite{BR}
\begin{gather}
E_n^{s}= - n \omega_c + s \sqrt{E_0^2 + \frac{2n v^2}{l_H^2}},  \,\, n= 1, 2, \dots, \, s=\pm, \notag \\
E_0 = - \frac{\omega_c}{2} -  \frac{g_L \mu_B H}{2} .
\label{eq:ee0}
\end{gather}
Here $l_H={1}/{\sqrt{|e|H}}$ and $\omega_c={|e|H}/{|m|}$ stands for the magnetic length and the cyclotron frequency, respectively. The corresponding wave functions in the Landau gauge, $\bm{A}=(-H y,0,0)$, reads
\begin{equation}
\psi_{n,s} =\frac{e^{ik_x x}}{\sqrt{L_x}}
\begin{pmatrix}
\alpha_{n,s} |n-1\rangle\\
\alpha_{n,-s} |n\rangle
\end{pmatrix} ,
\end{equation}
where $L_x$ denotes the size of the surface in the $x$ direction and $|n\rangle$ stands for standard states of Landau level problem. The coefficients $\alpha_{n,s}$ can be written as
\begin{equation}
\alpha_{n,s} = \frac{1}{\sqrt{1+D_n^2}}\begin{cases}
-i s D_n, & \quad s \sgn E_0 > 0 ,\\
1, & \quad s \sgn E_0 < 0 ,
\end{cases}
\end{equation}
where
\begin{equation}
D_n=\frac{\sqrt{2n}v/l_H}{|E_0|+\sqrt{E_0 ^2+2nv^2/l_H^2}} .
\end{equation}

To treat the hexagonal warping in the hamiltonian \eqref{eq:ham1:mag} as a perturbation, one
needs to evaluate matrix elements of the operator
\begin{equation}
V=\frac{\lambda}{2} \sum_{s=\pm} (k_{s}-eA_s)^3 \sigma_z \equiv \frac{\sqrt{2} \lambda}{l_H^3} \bigl ( {\hat a}^3+{\hat a}^{\dag 3}\bigr ) \sigma_z.
\end{equation}
Here the boson operators $\hat a$ and $\hat a^\dag$ are defined as follows
\begin{equation}
\hat a = \frac{l_H}{\sqrt{2}} \bigl (k_- - e A_- \bigr ), \qquad \hat a^\dag = \frac{l_H}{\sqrt{2}} \bigl (k_+ - e A_+ \bigr ) .
\end{equation}
The state $|n\rangle$ is the eigenstate of the operator ${\hat a}^\dag \hat a$, ${\hat a}^\dag \hat a | n\rangle = n | n\rangle$. Using the well-known matrix elements of the operators $\hat a$ and $\hat a^\dag$, we obtain the following results for the matrix elements
\begin{equation}
V_{n,n+3}^{s,s^\prime} =\frac{\sqrt{2}\lambda}{l_H^3}\bigl (\overline{\alpha_{n,s}} \alpha_{n+3,s^\prime}  \zeta_{n+2} -
\overline{\alpha_{n,-s}} \alpha_{n+3,-s^\prime} \zeta_{n+3}\bigr ) ,
\end{equation}
where $s,s^\prime=\pm$, 'bar' sign denotes complex conjugation,
and $\zeta_n=\sqrt{n(n-1)(n-2)}$ for $n\geqslant 0$. The other non-zero matrix elements can be obtained by complex conjugation. Hence, the second order correction to the eigenenergies \eqref{eq:ee0} due to the hexagonal warping is given as
\begin{equation}
\delta E_n ^{s, (2)}= -\sum_{s^\prime=\pm} \left (\frac{|V_{n+3,n}^{s^\prime,s}|^2}{E_{n+3}^{s^\prime}-E_n^s} +\frac{|V_{n-3,n}^{s^\prime,s}|^2}{E_{n-3}^{s^\prime}-E_n^s}\right ) .
\label{eq:pt1}
\end{equation}
For small values of $n$ (for low-lying Landau levels) the perturbation theory is applicable provided $\lambda/l_H^3 \ll \max\{\omega_c,v/l_H\}$. The second order correction $\delta E_n ^{s, (2)}$ grows with increase of $n$. Therefore, the perturbation theory breaks down at large $n$ if $\lambda$ is not sufficiently small. Denoting $\mathcal{X}={\lambda}/{v l_H^2}$ and $\mathcal{Y}=2|m| l_H v$ we find that the perturbation result \eqref{eq:pt1} is valid provided the following inequalities hold:
\begin{equation}
1 \gg \mathcal{X} n \begin{cases}
\mathcal{Y} \sqrt{n}, & \quad \mathcal{Y} \sqrt{n} \ll 1, \\
1, &  \quad 1 \ll \mathcal{Y} \sqrt{n} \ll  n, \\
\mathcal{Y}/ \sqrt{n}, & \quad n \ll \mathcal{Y} \sqrt{n} .
\end{cases}
\end{equation}

In addition the perturbation theory \eqref{eq:pt1} does not work near crossings of the unperturbed levels $E_n^+$ and $E_{n+3}^+$ that occur with varying magnetic field. To improve the perturbation theory near these degeneracy points we imply a unitary transformation of the hamiltonian which diagonalizes $2 \times 2$ matrix
\begin{equation}
A=\begin{pmatrix}
E_n ^+ & V_{n,n+3} ^{++}\\
V_{n,n+3} ^{++} & E_{n+3}^+
\end{pmatrix} .
\end{equation}
As usual, the eigenvalues of the matrix $A$
\begin{equation}
\Lambda_\pm=\frac{E_{n+3}^++E_n ^+}{2}\pm \frac{1}{2} \sqrt{(E_{n+3}^+-E_n ^+)^2+4 |V_{n,n+3}^{++}|^2} \,
\end{equation}
describe avoided crossing of levels $E_{n}^+$ and $E_{n+3}^+$ due to the matrix element $V_{n,n+3}^{++}$.
For a given $n$ we start from rewriting the hamiltonian \eqref{eq:ham1} in basis of the unperturbed states $\psi_{n,s}$:
\begin{equation}
\mathcal{H}=\begin{pmatrix}
A & B\\
B^\dagger & C
\end{pmatrix} .
\end{equation}

Here we introduce the following infinite block matrices
\begin{widetext}
 \begin{gather}
B=\begin{pmatrix}
V_{n,n+3}^{+-} & V_{n,n-3}^{++} & V_{n,n-3}^{+-} & 0 & 0 & 0 & \dots\\
0 & 0 & 0 & V_{n+3,n}^{+-} & V_{n+3,n+6}^{++} & V_{n+3,n+6}^{+-} & \dots
\end{pmatrix} ,
\notag \\
C=\begin{pmatrix}
E_{n+3}^- & 0 & 0 & V_{n+3,n}^{--} & V_{n+3,n+6}^{-+} & V_{n+3,n+6}^{--} & \dots\\
0 & E_{n-3}^+ & 0 & V_{n-3,n}^{+-} & 0 & 0 & \dots\\
0 & 0 & E_{n-3}^- & V_{n-3,n}^{--} & 0 & 0 & \dots \\
V_{n,n+3}^{--} & V_{n,n-3}^{-+} & V_{n,n-3}^{--} & E_n ^- & 0 & 0 & \dots\\
V_{n+6,n+3}^{+-} & 0 & 0 & 0 & E_{n+6}^+ & 0 & \dots\\
V_{n+6,n+3}^{--} & 0 & 0 & 0 & 0 & E_{n+6}^- & \dots\\
\dots & \dots & \dots & \dots& \dots& \dots& \dots
\end{pmatrix} .
\end{gather}
\end{widetext}
The unitary transformation diagonalizing the matrix $A$ is as follows
\begin{equation}
U=\begin{pmatrix}
u & 0\\
0 & 1
\end{pmatrix} , \qquad u=\begin{pmatrix}
\frac{1}{\sqrt{1+\gamma_+ ^2}} & \frac{1}{\sqrt{1+\gamma_- ^2}}\\
\frac{\gamma_+}{\sqrt{1+\gamma_+ ^2}} & \frac{\gamma_-}{\sqrt{1+\gamma_- ^2}}
\end{pmatrix} ,
\end{equation}
\begin{equation}
\gamma_\pm = \frac{ E_{n+3}^+ -E_n ^+ \pm \sqrt{(E_n ^+ - E_{n+3}^+)^2+4|V_{n,n+3}^{+-}|^2}}{2V_{n,n+3}^{++}} .
\end{equation}

\begin{figure}[t]
\centerline{\includegraphics[width=0.35\textwidth]{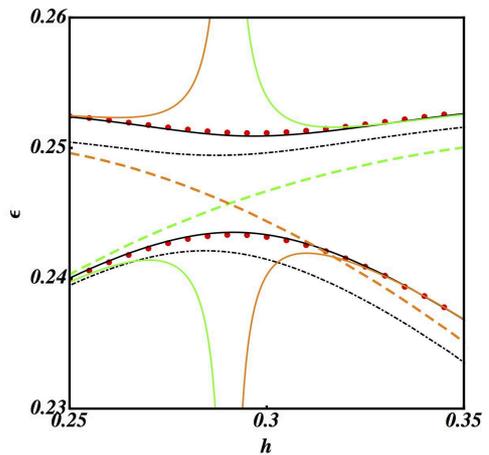}}
\caption{(Color online) The dependence of dimensionless unperturbed energies $E_{4}^+/\Delta$ and $E_7^+/\Delta$ on dimensionless magnetic field $h = 4\pi v^2/(l_H \Delta)^2$ (dashed orange and green curves) near their crossing point. The thin solid orange and green curves illustrate the results of the standard perturbation theory (see Eq. \eqref{eq:pt1}). The dot-dashed black curves are the eigenvalues $\Lambda_\pm$.
The thick solid black curves are the result of modified perturbation theory (see Eq. \eqref{eq:pt2}). Red points represent the results of numerical diagonalization of the truncated hamiltonian with $2000$ levels. The dimensionless parameter of the hexagonal warping is $\alpha=0.1$ and $g_L=0$.}
\label{Fig:Fig6}
\end{figure}

Now taking into account the matrix elements (given by $u^\dagger B$) connecting levels $\Lambda_\pm$ with the other levels within the second order perturbation theory we find the following results for energies corresponding to the unperturbed energies $E_{n}^+$ and $E_{n+3}^+$:
\begin{align}
E_\pm   = \Lambda_\pm & + \frac{1}{1+\gamma_\pm^2} \Biggl ( \frac{|V_{n,n+3}^{+-}|^2}{\Lambda_\pm-E_{n+3}^-} +\frac{|V_{n,n-3}^{++}|^2}{\Lambda_\pm-E_{n-3}^+}
\notag \\
& +\frac{|V_{n,n-3}^{+-}|^2}{\Lambda_\pm-E_{n-3}^-}+\frac{|\gamma_\pm V_{n+3,n}^{+-}|^2}{\Lambda_\pm-E_{n}^-}
\notag \\
& +\frac{|\gamma_\pm V_{n+3,n+6}^{++}|^2}{\Lambda_\pm-E_{n+6}^+}+\frac{|\gamma_\pm V_{n+3,n+6}^{+-}|^2}{\Lambda_\pm-E_{n+6}^-}\Biggr ) .
\label{eq:pt2}
\end{align}
This result is free from fictitious divergence at the point $E_{n}^+=E_{n+3}^+$ produced within the standard perturbation theory, Eq. \eqref{eq:pt1}. Away from the crossing point, the result \eqref{eq:pt2} transforms into the result \eqref{eq:pt1}. We illustrate the result \eqref{eq:pt2} of the modified perturbation theory, which is essentially the correct choice of wave functions for the zero-order approximation, in Fig. \ref{Fig:Fig6} for the crossing of the unperturbed levels $E_{4}^+$ and $E_7^+$. As one can see from Fig. \ref{Fig:Fig6} the expressions \eqref{eq:pt2} smoothly interpolate the results of the standard second-order perturbation theory, Eq. \eqref{eq:pt1}, before and after the degeneracy point. Even in the close vicinity of the crossing point, the energies $E_\pm$ are different from the eigenvalues $\Lambda_\pm$ of matrix $A$, i.e., the transitions to the other levels are important. The energy levels found from Eq. \eqref{eq:pt2}  are in good agreement with numerical diagonalization of the hamiltonian \eqref{eq:ham1:mag}.

\section{Landau levels in the WKB approximation \label{sec3}}

To study the structure of Landau levels at higher energies we use the WKB approach. \cite{mag} We employ the Bohr-Sommerfeld quantization condition:
\begin{equation}
S(E)=2\pi l_H ^{-2}\bigl ( n+\delta(E)\bigr ) ,
\label{eq:BS}
\end{equation}
where $S(E)$ denotes the area bounded by a curve of the constant energy $E$ in the momentum space in the absence of magnetic field, $n$ is an integer number, $\delta(E)$ involves the information on the number of turning points of a quasiclassical electron orbit and the Berry phase. \cite{Falkovsky} Typically the function $\delta(E)$ is of the order unity. Since we are interested in Landau levels with $n\gg 1$ we omit $\delta(E)$ below. Also we neglect the Zeeman splitting  assuming that $g$-factor is not strongly enhanced in comparison with its band value.

The area $S(E)$ can be expressed through the density of states without magnetic field. As it follows from results of Sec. \ref{sec1},  for some values of $\epsilon$ and $\alpha$ there are several disconnected regions enclosed by constant-energy curve. In this case, the quasiclassical quantization condition \eqref{eq:BS} has to be applied to each disconnected area separately. For energies in the interval $0<\epsilon<\min\{\epsilon_+(\alpha),1/4\}$ (see regions A1 and A4 in Fig. \ref{Fig:Fig1}b) there is one snowflakelike region including the $\bar{\Gamma}$ point and six outermost regions of infinite area (see Fig. \ref{Fig:Fig3}). The area of the central region can be written as
\begin{equation}
S_1=\frac{\Delta^2}{2v^2}\left [2\pi c_1 +12 \int\limits_{c_1}^{x_1}dx \, \mathcal{G}(x,\epsilon,\alpha) \right ] ,
\end{equation}
where we introduce the function
\begin{equation}
\mathcal{G}(x,\epsilon,\alpha)=\frac13\arccos\frac{-\sqrt{(x+\epsilon)^2-x}}{\sqrt{\alpha x^3}}-\frac{\pi}{6} .
\end{equation}
It can be shown that
\begin{equation}
\frac{\partial S_1}{\partial \epsilon}=4\pi^2\Delta g_1(\epsilon) , \qquad g_1(\epsilon)=\frac{\Delta}{2\pi v^2}F_1(\epsilon,\alpha) .
\label{eq:s1}
\end{equation}
The function $g_1(\epsilon)$ provides the contribution to the density of states $g(\epsilon)$ from the states in this snowflakelike central region.  The area of each among six outermost regions is given as
\begin{equation}
S_5=\frac{\Delta^2}{v^2} \int\limits_{x_2}^{+\infty} dx \, \mathcal{G}(x,\epsilon,\alpha) .
\label{eq:ss2}
\end{equation}
Again this area can be related to the corresponding contribution to the density of states:
\begin{equation}
\frac{\partial S_5}{\partial \epsilon}=\frac{2\pi^2\Delta}{3} g_5(\epsilon) , \qquad g_5(\epsilon)=\frac{\Delta}{2\pi v^2}F_5(\epsilon,\alpha) .
\end{equation}
Since the integral in Eq. \eqref{eq:ss2} diverges at the upper limit, it is convenient to rewrite Eq. \eqref{eq:ss2} as follows:
\begin{equation}
S_5(\epsilon)=\frac{1}{6}S(0)+\frac{\pi\Delta^2}{3v^2} \int\limits_0^\epsilon d\epsilon^\prime F_5(\epsilon^\prime,\alpha)  .
\end{equation}
Here $S(0)$ is the total area enclosed by the constant energy curve $\epsilon=0$. We note that in the framework of the hamiltonian \eqref{eq:ham1} the area $S(0)$ is infinite. It becomes finite if one takes into account, for example, the next order in $k^2$ correction to the mass $m$. Within the quasiclassical approximation the Bohr-Sommerfeld quantization condition \eqref{eq:BS} for $S_5(\epsilon)$ results in
sixfold degenerate levels. The quantum tunneling (magnetic breakdown) removes this degeneracy. \cite{mag}

In the case $\max\{0,\epsilon_+(\alpha)\}< \epsilon < 1/4$ (see region A3 in Fig. \ref{Fig:Fig1}b), there are two  disconnected parts of the area (see Fig. \ref{Fig:Fig3}). The area of the innermost part is given by Eq. \eqref{eq:s1}, whereas the area of the outermost part reads
\begin{align}
S_{2,3}
& = \frac{6\Delta^2}{v^2}\left ( \int\limits_{x_2}^{c_2} dx \, \mathcal{G}(x,\epsilon,\alpha) +\int\limits_{c_3}^{+\infty} dx \, \mathcal{G}(x,\epsilon,\alpha)\right ) \notag \\
& + \frac{\Delta^2}{2v^2}(2\pi c_3-2\pi c_2) .
\end{align}
Again, we find
\begin{equation}
\frac{\partial S_{2,3}}{\partial \epsilon}=4\pi^2\Delta g_{2,3}(\epsilon) ,
\,
g_{2,3}(\epsilon)=\frac{\Delta}{2\pi v^2}\bigl (F_2(\epsilon,\alpha)+ F_3(\epsilon,\alpha)\bigr ) .
\end{equation}
It is convenient to rewrite $S_{2,3}$ as follows:
\begin{equation}
S_{2,3}=S(0)+\frac{2\pi\Delta^2}{v^2} \int\limits_0^\epsilon d\epsilon^\prime \Bigl [ F_2(\epsilon^\prime,\alpha)+F_3(\epsilon^\prime,\alpha)\Bigr ]   .
\end{equation}
In the other case $1/4< \epsilon$ (see Fig. \ref{Fig:Fig3}) there is always one connected region whose area can be written as
\begin{equation}
S(\epsilon) = S(0) + 4\pi^2\Delta \int\limits_0^\epsilon d\epsilon^\prime g(\epsilon^\prime) .
\end{equation}
For $\epsilon<0$ the area can be found using the following relation:
\begin{equation}
\frac{\partial S}{\partial \epsilon}=4\pi^2\Delta \bigl (g_5(\epsilon)-g_1(\epsilon)\bigr ) .
\end{equation}

\begin{figure}[t]
\centerline{(a)\includegraphics[width=0.22\textwidth]{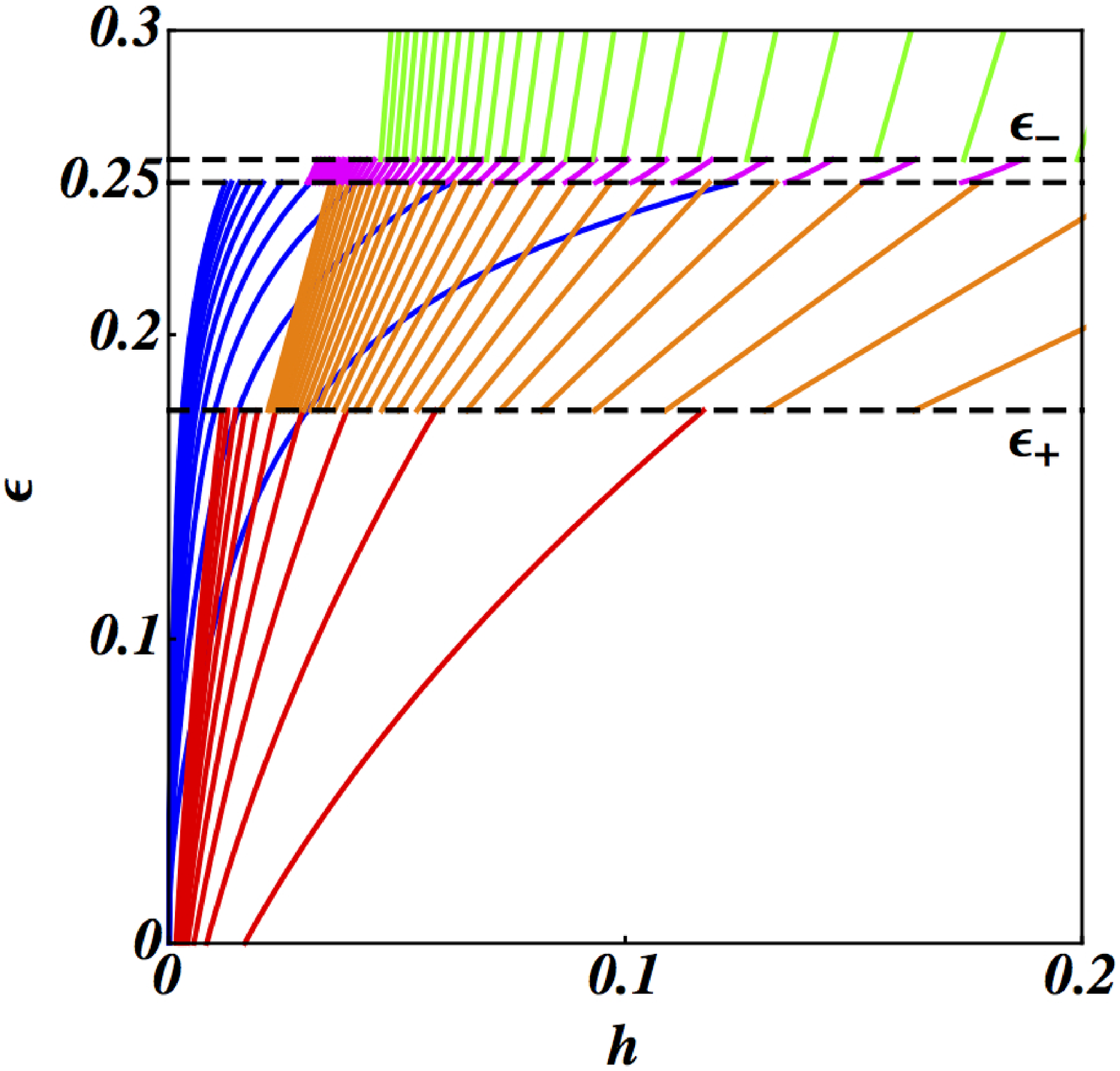}\quad (b)\includegraphics[width=0.22\textwidth]{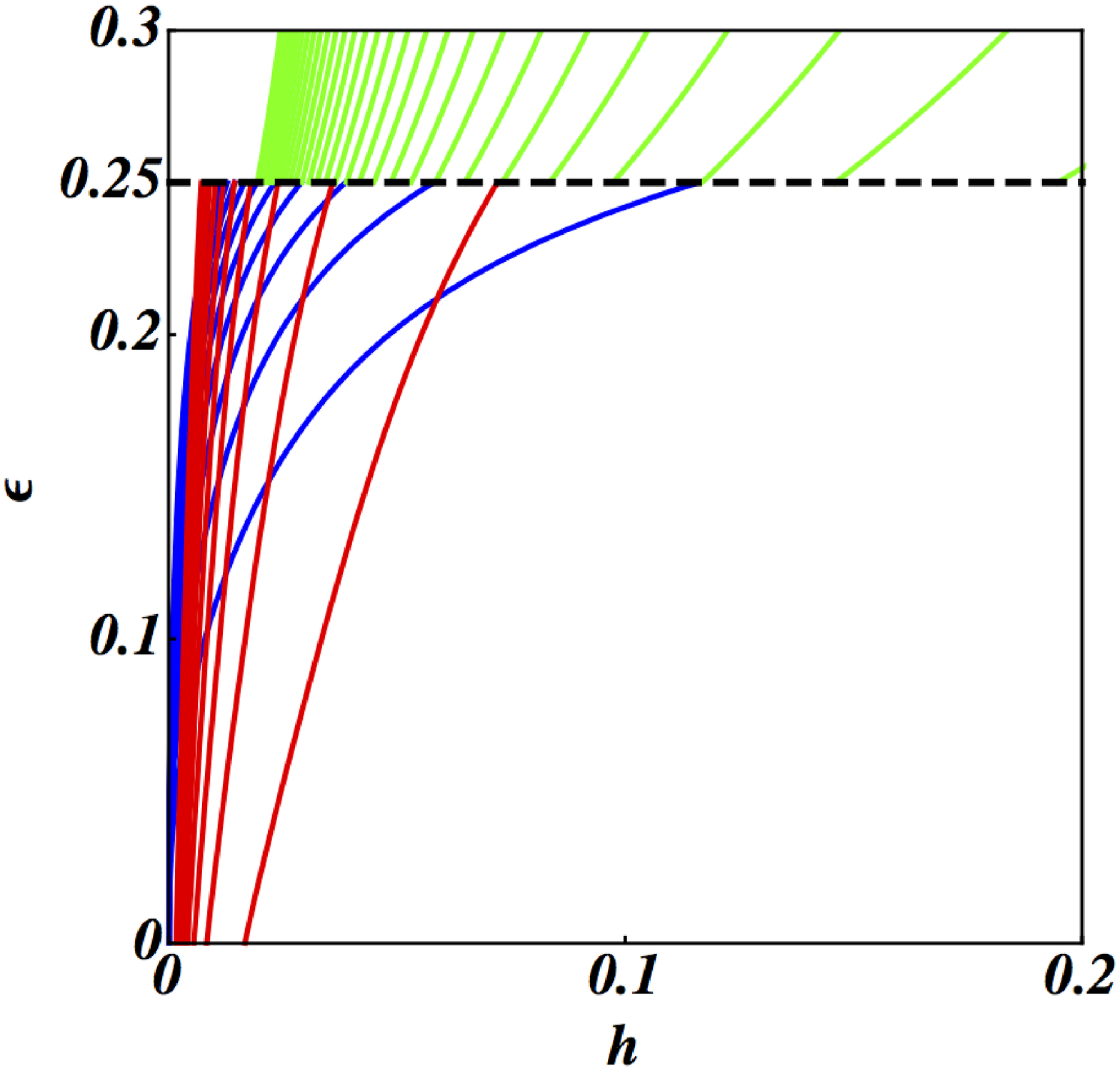}}
\caption{(Color online) The structure of Landau levels in the WKB approximation (each 10th level is shown) for (a) $\alpha=0.4$ and (b) $\alpha=2$. Blue curves denote the levels due to the central snowflakelike area $S_1$. Red curves are sixfold degenerate levels. Orange curves correspond to  levels due to the area obtained after consolidation of six disconnected outermost pieces. Magenta curves  are the levels corresponding to unified area but with six empty spots. Green curves denote the levels for the case when the empty spots disappear. The total area at $\epsilon=0$ is chosen to be equal to $S(0) = \Delta^2/(2v^2)$.
}
\label{Fig:Fig7}
\end{figure}

\begin{figure}[t]
\centerline{(a)\includegraphics[width=0.22\textwidth]{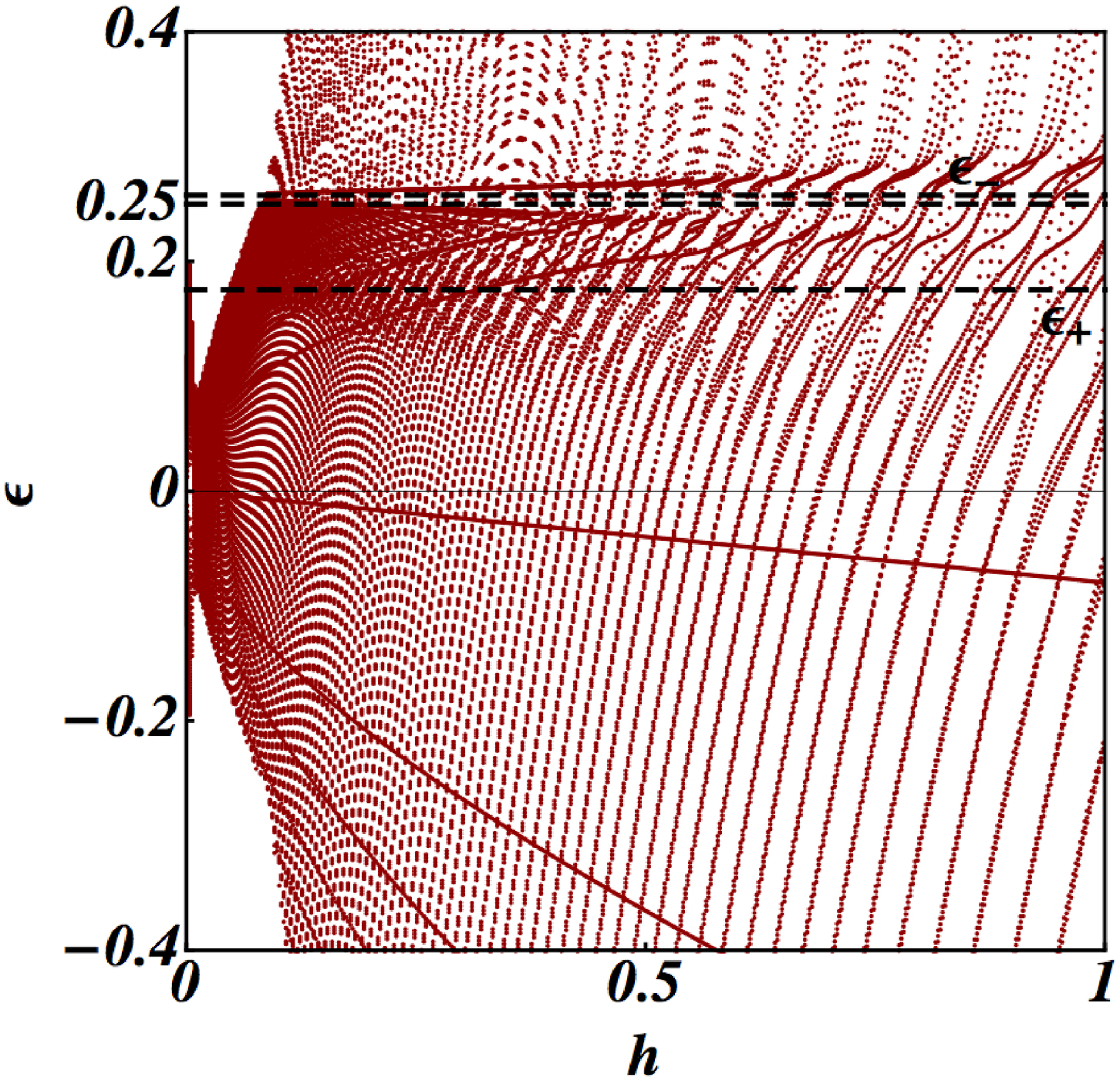}\quad (b)\includegraphics[width=0.22\textwidth]{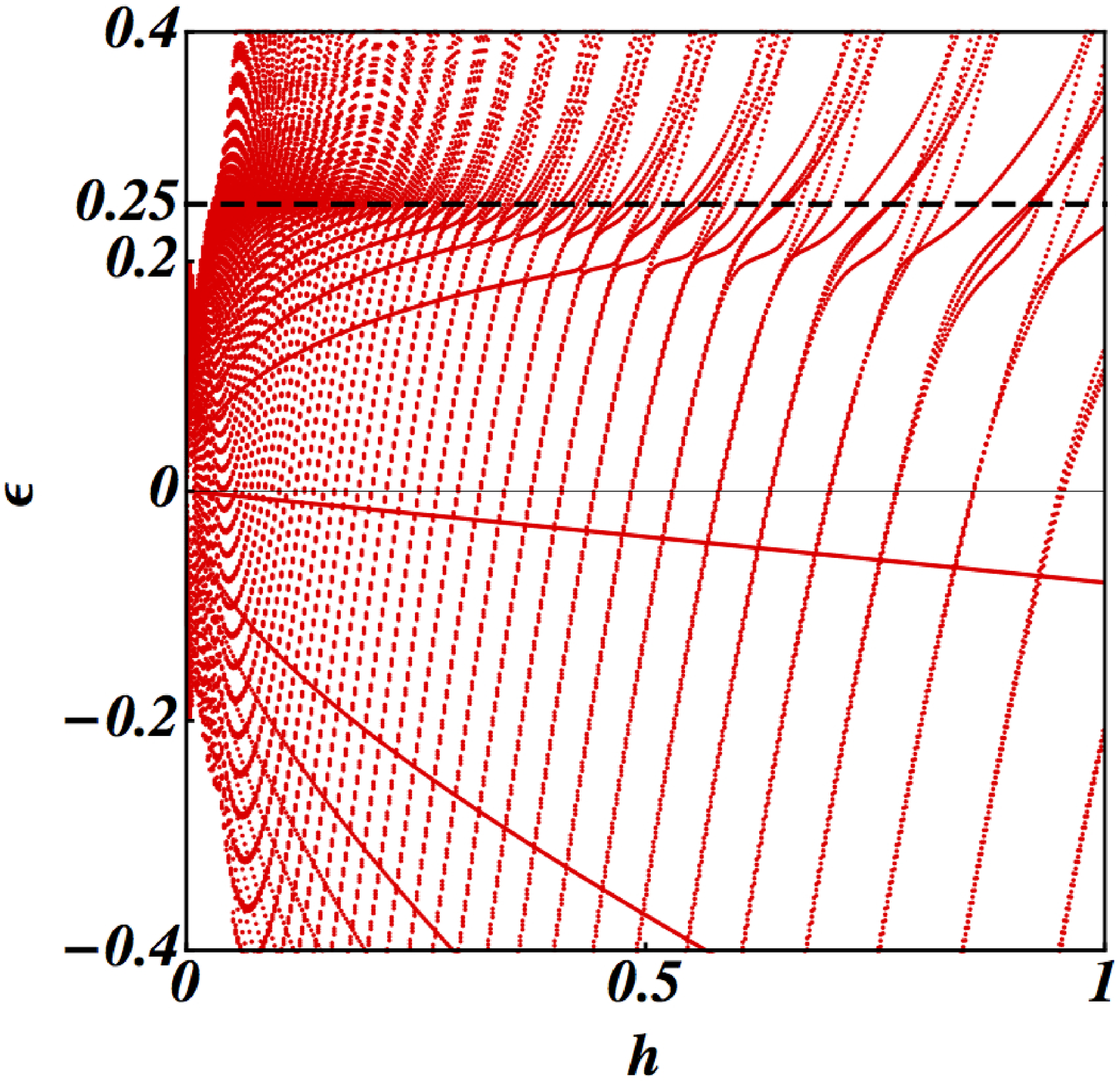}}
\caption{(Color online) The structure of Landau levels from numerical diagonalization of the truncated hamiltonian with $2000$ levels for (a) $\alpha=0.4$ and (b) $\alpha=2$.
}
\label{Fig:Fig8}
\end{figure}

The structure of Landau levels undergoes changes near such singularities of the zero-field density of states which are related to the change of number of connected parts of the area enclosed by the constant-energy curve.

For $\alpha<\alpha_0$ the sixfold degenerate levels transform into non-degenerate levels at $\epsilon=\epsilon_+(\alpha)$. Using Eq. \eqref{eq:a2} we can estimate the change in the level spacing at $\epsilon=\epsilon_+(\alpha)$. We find
\begin{equation}
\frac{d\epsilon}{dn} = \frac{h}{4\Delta_+ \ln (1/|\epsilon-\epsilon_+|)}
\begin{cases}
6 , & \quad \epsilon_+-\epsilon\ll 1 ,\\
1, & \quad \epsilon-\epsilon_+ \ll 1,
\end{cases}
\end{equation}
where $h=4\pi v^2/(l_H\Delta)^2$ stands for dimensionless magnetic field. Thus the sixfold degenerate levels (corresponding to six disconnected pieces) are 6 times sparser than the levels after the disconnected pieces merged together. Also the slope of the sixfold degenerate levels with respect to magnetic field is $6$ times larger than the slope of levels after consolidation of the disconnected pieces. The levels corresponding to the area $S_1$ are continuous at $\epsilon=\epsilon_+(\alpha)$. However at $\epsilon=1/4$ the area $S_1$ merges with the area $S_{2,3}$. Using Eq. \eqref{eq:a1} we can estimate the level spacing before and after consolidation:
\begin{equation}
\frac{d\epsilon}{dn} = \frac{h \sqrt{\alpha}}{16 \ln (1/|\epsilon-1/4|)}
\begin{cases}
2 , & \quad 1/4-\epsilon\ll 1 ,\\
1, & \quad \epsilon-1/4 \ll 1 .
\end{cases}
\end{equation}
Each of Landau levels corresponding to the areas  $S_1$  and $S_{2,3}$ are twice sparser than the levels after consolidation. Also the slope of these levels at $\epsilon=1/4$ becomes 2 times smaller.

For $\alpha>\alpha_0$ Landau levels undergo reconstruction at $\epsilon=1/4$ only. At $1/4-\epsilon\ll 1$
there are two sets of levels: the sixfold degenerate ($\epsilon_{2,3}$) and nondegenerate ($\epsilon_1$) ones with the level spacings
\begin{equation}
\frac{d\epsilon_1}{dn} = \frac{h \sqrt{\alpha}}{8\ln [1/(1/4-\epsilon)]},
\quad  \frac{d\epsilon_{2,3}}{dn} = \frac{3 h \sqrt{\alpha}}{4\ln [1/(1/4-\epsilon)]} .
\end{equation}
The sixfold degenerate levels are 6 times sparser  and steeper than the levels after the disconnected pieces merged together. At $\epsilon > 1/4$ there is only single set of Landau levels with the spacing:
 \begin{equation}
\frac{d\epsilon}{dn} = \frac{h \sqrt{\alpha}}{16 \ln [1/(\epsilon-1/4)]} .
\end{equation}
These levels are 2 times rarer and smoother than $\epsilon_1$ levels.

We illustrate  transformations of Landau levels  discussed above in Fig. \ref{Fig:Fig7} for two values of the dimensionless parameter of the hexagonal warping, $\alpha=0.4$ and $\alpha=2$. There are several interesting features due to the hexagonal warping in structure of the Landau levels. At first, the hexagonal warping leads to existence of the sixfold degenerate levels (red curves in Fig. \ref{Fig:Fig7}) within WKB approximation for $\epsilon < \min\{\epsilon_+,1/4\}$. The account of quantum tunneling (magnetic breakdown) should remove this degeneracy. Secondly, due to the hexagonal warping there exist levels (green curves in Fig. \ref{Fig:Fig7}) with energies well above $\Delta/4$ which is not possible in the case $\alpha=0$. However in the WKB approximation it is not clear how the Landau levels at $\alpha=0$ transform to produce levels with energies above $\Delta/4$ in  the case of $\alpha>0$. Therefore, we compare the results of the WKB approximation with Landau levels obtained by numerical diagonalization of hamiltonian \eqref{eq:ham1:mag} truncated by $2000$ levels. As one can see from Fig. \ref{Fig:Fig8} the numerical results are in qualitative agreement with the quasiclassical treatment.

\section{Experimental results \label{sec4}}

The compound of Bi$_2$Te$_3$ represents the topological insulator where the singularities of the TDOSS due to the finite curvature and hexagonal warping can be observed most probably. In typical case of positive mass and not very small hexagonal warping all singularities discussed in this paper are situated below the Dirac point. Thus they can be hidden or even destroyed by bulk contributions. The case of negative mass, as in Bi$_2$Te$_3$, is special since singularities of TDOSS are situated above the Dirac point. 

\begin{figure}[t]
\centerline{\includegraphics[width=0.55\textwidth]{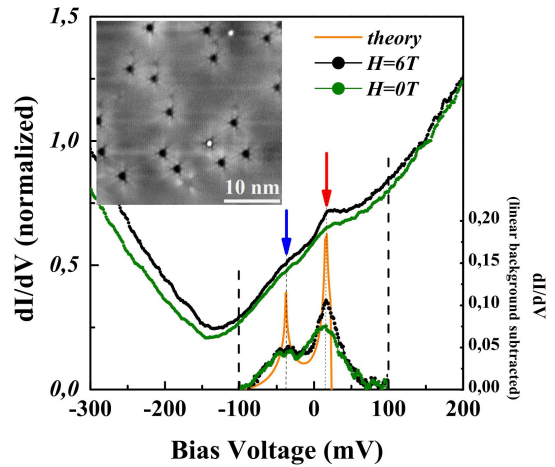}}
\caption{(Color online) Two local tunneling $dI/dV(V)$ spectra of Bi$_2$Te$_3$ at the $0.3$ K:
green curve -- at zero magnetic field, black curve -- at 6 T. Color arrows mark positions of step-like cusps (see in the text). Top inset: Scanning topography image of the studied surface with several atomic defects. Bottom inset: 
The same $dI/dV(V)$ spectra with a linear background subtracted. The orange curve is theoretical TDOSS for $\alpha=0.44$ (see text).}
\label{Fig:Fig:Exp}
\end{figure}


We indeed observed singularities in the TDOSS by providing scanning tunneling microscopy/spectroscopy experiment on in-vacuum cleaved surface of Bi$_2$Te$_3$.  
The sample we used in this work has been recently characterized by ARPES. \cite{Scholz2013}
In Fig. \ref{Fig:Fig:Exp} we present a scanning tunneling microscopy image of the studied surface. The surface is atomically flat; yet several individual atomic defects are visible, appearing as three-fold stars. The two presented curves correspond to the local tunneling conductance dI/dV(V) measured at zero magnetic field (green curve) and at 6T perpendicular field (black curve). At zero field a step-like spectroscopic feature is observed at around $+25$ mV (marked by red arrow), i.e. above the Fermi level, as theoretically expected for this material (such step-like cusps are also present in the data of other tunneling experiments \cite{ZA1,ZA2}). Moreover, we found that in strong magnetic field this step-like feature transforms into a distinct maximum (black curve). Another yet weaker step-like feature is observed at around $-50$ mV (marked by blue arrow). 
We conjecture that both step-like features are related to singularities in the TDOSS at $E = \Delta/4$ and $E = \Delta \epsilon_+$. The estimated position of the Dirac point is around $-240$ mV, in agreement with Ref. [\onlinecite{Scholz2013}]. From the distance between the Dirac point and the step-like cusp at around $+25$ mV we find $\Delta \approx 1$ V. The fitting of the spectral position of the double-peaked structure at around $-50$ mV and $+25$ mV is obtained for $\alpha = 0.44$ (see the orange curve in Fig. \ref{Fig:Fig:Exp}). The theoretical curve describes reasonably well the position of the observed peaks, yet it fails in reproducing their significantly larger spectral width. We notice that numerous atomic defects present at the surface (see Fig. \ref{Fig:Fig:Exp}) may affect the local Fermi level, and lead to a smearing of logarithmic singularity in the experimental tunneling spectra. Taking into account this effect theoretically is a complex task, well beyond the main goal of the present work.
A more detailed investigation of this effect is needed;š it requires a full mapping of the local density of states, and will be subject of a separate report.

\section{Discussions and conclusions \label{sec5}}

Using recent results of spin and angle resolved photoemission spectroscopy \cite{NS} we estimate the parameters relevant for the model considered above for two topological insulators Bi$_2$Te$_3$ and Bi$_2$Se$_3$. We note that they differ by the sign of the effective mass $m$. It is negative for Bi$_2$Te$_3$ and positive for Bi$_2$Se$_3$. Estimates for parameters of the model extracted from Ref. [\onlinecite{NS}] are summarized in Table \ref{Tab2}. We emphasize that although the energy scales $\Delta$ and $E_0$ are of the same order for both topological insulators, the dimensionless parameter $\alpha$ characterizing the strength of the hexagonal warping differs by more than 50 times. From our experiment we obtain estimate for $\Delta$ which is close to the value reported in Ref.  [\onlinecite{NS}]  for Bi$_2$Te$_3$. However, the two step-like cusps structure revealed in our data suggests that the hexagonal warping in studied Bi$_2$Te$_3$ sample is significantly weaker  than reported in Ref.  [\onlinecite{NS}], $\alpha=0.44$ instead of $\alpha=22$.

The logarithmic singularity in the TDOSS at $E=\Delta/4$ corresponds to consolidation of snowflakelike central region and the six outermost disconnected regions. It occurs in certain directions of the momentum space, e.g. at the angle $\theta=\pi/6$. The condition $E_+(k_0,\pi/6) = \Delta/4$ is solved by the momentum $k_0=\Delta/(2v)$. According to estimates in Table \ref{Tab2} it is much smaller than a size of the surface Brillouin zone which is of the order of 1 \AA$^{-1}$. Also we note that for such momentum the ratio of  the hexagonal warping term to the linear in momentum term is of the order of $\lambda k_0^2/v = \sqrt{\alpha}/4$. It indicates that for $\alpha/16 \ll 1$ the singularity occurs in the regime where the hexagonal warping is a small correction to the linear in $k$ dispersion. These estimates are in favor of using the hamiltonian \eqref{eq:ham1}, which was derived near the $\bar{\Gamma}$ point, to describe the singularity in TDOSS at $E=\Delta/4$.

\begin{table}[t]
\caption{Estimates for parameters of the model extracted from Ref. [\onlinecite{NS}] (see text).}
\begin{tabular}{c|ccccc}
& $\Delta$, eV & $E_0$, eV & $\alpha$ & $k_0$, \AA$^{-1}$ & $h/H$, T$^{-1}$\\
\hline
Bi$_2$Te$_3$ & 1.1  & 0.51 & 22 & 0.14 & 2.2 $\cdot$ 10$^{-3}$\\
Bi$_2$Se$_3$ & 0.34 & 0.43 & 0.4 & 0.08 & 7.0 $\cdot$ 10$^{-3}$
\end{tabular}
\label{Tab2}
\end{table}

Finally, we stress the smallness of dimensionless magnetic field $h$ for both Bi$_2$Te$_3$ and Bi$_2$Se$_3$ (see Table \ref{Tab2}). It implies the smallness of the parameter $\omega_c l_H/v = \sqrt{h/\pi}$.
Validity of the perturbation theory for Landau levels with small level index is controlled by the parameter $\sqrt{\alpha} h/(4\pi)$. Therefore, for moderate values of $\alpha$ low-lying Landau levels are not significantly affected by presence of the finite curvature and hexagonal warping and, thus, scale as $\sqrt{H}$. Such scaling for Landau levels near the Dirac point was recently observed from oscillations in the tunneling conductance of 
Bi$_2$Se$_3$, \cite{Hanaguri2010} of Sb$_2$Te$_3$, \cite{YY} from microwave spectroscopy in Bi$_2$Te$_3$, \cite{Wolos2012} and from magneto-infrared spectroscopy in Bi$_{0.91}$Sb$_{0.09}$. \cite{SP} The effect of the hexagonal warping is most pronounced near the degeneracy points of the unperturbed Landau levels. For a given $h \ll 1$, the degeneracy point corresponds to Landau levels with $n_h \sim \pi/(2h) \gg 1$ and energies of the order of $\Delta/4$. The hexagonal warping leads to avoided crossing of Landau levels $E_{n_h}^+$ and $E_{n_h+3}^+$ with the typical distance between them of the order of  $\delta_h \sim \sqrt{\alpha/2} \Delta h/(8\pi)$. Additional signature of the hexagonal warping is the existence of oscillations in the tunneling conductance in magnetic field at energies above $\Delta/4$. In the case of Bi$_2$Te$_3$ for magnetic field $H = 10$ T we can estimate $n_h\approx 70$ and $\delta_h \sim 3$ meV. 
We expect that future tunneling experiments on topological insulators with warped electronic spectra will indeed reveal the predicted complex structure of LLs and their unusual evolution in magnetic field.

To summarize, we computed the tunneling density of surface states $g(E)$ in a 3D topological insulator in the presence of the hexagonal warping and finite curvature. We found that the hexagonal warping transforms the square-root van Hove singularity of $g(E)$ into the logarithmic one. With increase of the hexagonal warping the singularity becomes weaker. For values of the hexagonal warping $\lambda \lesssim 0.18/m^2 v$ the tunneling density of states has the additional logarithmic singularity and the jump. Their positions and amplitudes depend on $\lambda$.
In the case of Bi$_2$Te$_3$ we experimentally observe two step-like cusps in the tunneling density of states at around $-50$ mV and $+25$ mV. They are identified as fingerprints of such logarithmic singularities situated, as expected for this material, above the Dirac point. In the presence of the perpendicular magnetic field we analyzed structure of the Landau levels within the perturbation theory in the hexagonal warping and in the WKB approximation. We obtained that the hexagonal warping removes degeneracies of the Landau levels and changes drastically their behavior with the magnetic field.

\begin{acknowledgments}

The theoretical part of the research was funded by the Russian Science Foundation under the grant No. 14-12-00898. The experimental part of the research was supported by Ministry of Education and Science of the Russian Federation, grant No. 14Y.26.31.0007 and through European COST action network.

\end{acknowledgments}


\begin{thebibliography}{100}

\bibitem{HK} M. Z. Hasan and C. L. Kane, Rev. Mod. Phys. {\bf 82}, 3045 (2010).
\bibitem{QZ} X.-L. Qi and S.-C. Zhang, Rev. Mod. Phys. {\bf 83}, 1057 (2011).
\bibitem{Ando2013} Y. Ando, J. Phys. Soc. J. {\bf 82}, 102001 (2013).
\bibitem{F} L. Fu, Phys. Rev. Lett. {\bf 103}, 266801 (2009).
\bibitem{LQ} C.-X. Liu, X.-L. Qi, H. J. Zhang, X. Dai, Z. Fang, and S.-C. Zhang, Phys. Rev. B {\bf 82}, 045122 (2010).

\bibitem{Chen2009} Y. Chen, J. G. Analytis, J.-H. Chu, Z. K. Liu, S.-K. Mo, X. L. Qi, H. J. Zhang, D.H. Lu,
X. Dai, Z. Fang, S.-C. Zhang, I. R. Fisher, Z. Hussain, and Z.-X. Shen, Science {\bf 325}, 178 (2009).

\bibitem{ZA1} Zh. Alpichshev, J. G. Analytis, J.-H. Chu, I. R. Fisher, and A. Kapitulnik, Phys. Rev. B {\bf 84}, 041104 (2011).

\bibitem{Kuroda2010} K. Kuroda, M. Arita, K. Miyamoto, M. Ye, J. Jiang, A. Kimura, E.E. Krasovskii, E.V. Chulkov, H. Iwasawa, T. Okuda, K. Shimada, Y. Ueda, H. Namatame, and M. Taniguchi, Phys. Rev. Lett. {\bf 105}, 076802  (2010).

\bibitem{NS} M. Nomura, S. Souma, A. Takayama, T. Sato, T. Takahashi, K. Eto, K. Segawa, and Y. Ando, Phys. Rev. B {\bf 89}, 045134 (2014).

\bibitem{Wang2011} C. M. Wang and F. J. Yu, Phys. Rev. B {\bf 84}, 155440 (2011).

\bibitem{Xiao2013} X. Xiao and W. Wen, Phys. Rev. B {\bf 88}, 045442 (2013).

\bibitem{Smirnov2013} S. Smirnov, Phys. Rev. B {\bf 88}, 205301 (2013).

\bibitem{Fu2013} Z.-G. Fu, F. Zheng, Z. Wang, and P. Zhang, Prog. Theor. Exp. Phys. 103I01 (2013).

\bibitem{Urazhdin} S. Urazhdin, D. Bilc, S. D. Mahanti, S. H. Tessmer, Th. Kyratsi, and M. G. Kanazidis, Phys. Rev. B {\bf 69}, 085313 (2004).

\bibitem{ZA2} Zh. Alpichshev, J. G. Analytis, J.-H. Chu, I. R. Fisher, Y. L. Chen, Z. X. Shen, A. Fang, and A. Kapitulnik, Phys. Rev. Lett. {\bf 104}, 016401 (2010).

\bibitem{SO} P. Sessi, M. M. Otrokov, T. Bathon, M. G. Vergniory, S. S. Tsirkin, K. A. Kokh, O. E. Tereshchenko, E. V. Chulkov, M. Bode, Phys. Rev. B {\bf 88}, 161407(R) (2013).

\bibitem{Hanaguri2010} T. Hanaguri, K. Igarashi, M. Kawamura, H. Takagi, and T. Sasagawa, Phys. Rev. B {\bf 82}, 081305(R) (2010).

\bibitem{tZhang2013} T. Zhang, N. Levy, J. Ha, Y. Kuk, and J. A. Stroscio, Phys. Rev. B {\bf 87}, 115410 (2013).

\bibitem{YY} Y. Jiang, Y. Wang, M. Chen, Z. Li, C. Song, K. He, L. Wang, X. Chen, X. Ma, Q.-K. Xue1, Phys. Rev. Lett. {\bf 108}, 016401 (2012).

\bibitem{YSFu2014} Y.-S. Fu, M. Kawamura,  K. Igarashi, H. Takagi, T. Hanaguri, and T. Sasagawa, arxiv:1408.0873 (unpublished).

\bibitem{Zaitsev2014} A. Yu. Dmitriev, N. I. Fedotov, V. F. Nasretdinova, and S. V. Zaitsev-Zotov, arxiv:1408.4991 (unpublished).




\bibitem{Saha2011} K. Saha, S. Das, K. Sengupta, and D. Sen, Phys. Rev. B {\bf 84}, 165439 (2011).

\bibitem{Schwab2012} P. Schwab and M. Dzierzawa, Phys. Rev. B {\bf 85}, 155403 (2012).

\bibitem{Vazifeh2012} M. M. Vazifeh and M. Franz, Phys. Rev. B {\bf 86}, 045451 (2012).

\bibitem{BR} Yu. A. Bychkov and E. I. Rashba,  JETP Lett. {\bf 39}, 78 (1984); J. Phys. C: Sol. State Phys. {\bf 17}, 6039 (1984).



\bibitem{Silvestrov2012} P. G. Silvestrov, P. W. Brouwer, and E. G. Mishchenko, Phys. Rev. B {\bf 86}, 075302 (2012).

\bibitem{fZhang2012} F. Zhang, C. L. Kane, E. J. Mele, Phys. Rev. B {\bf 86}, 081303(R) (2012).


\bibitem{Basak2011} S. Basak, H. Lin, L.A. Wray,  S.-Y. Xu, L. Fu, M. Z. Hasan, and A. Bansil, Phys. Rev. B {\bf 84}, 121401(R) (2011).



\bibitem{YH} Z. Yang and J. H. Han, Phys. Rev. B {\bf 83}, 045415 (2011).


\bibitem{mag} for a review, see e.g. I. M. Lifshitz, M. Y. Azbel, M. I. Kaganov, {\it Electronic theory of metals}, M.: Nauka (1971).

\bibitem{Falkovsky} L. A. Falkovsky, JETP {\bf 49}, 609 (1965); A.Yu. Ozerin, L.A. Falkovsky, Phys. Rev. B {\bf 85}, 205143 (2012).

\bibitem{Scholz2013} M. R. Scholz, J. S\'anchez-Barriga, J. Braun,  D. Marchenko, A. Varykhalov, M. Lindroos, Y. J. Wang, H. Lin, A. Bansil, J. Min\'ar, H. Ebert, A. Volykhov, L. V. Yashina, and O. Rader, Phys. Rev. Lett. {\bf 110}, 216801 (2013).

\bibitem{Wolos2012} A. Wolos, S. Szyszko,  A. Drabinska, M. Kaminska, S. G. Strzelecka, A. Hruban, A. Materna, and M. Piersa, Phys. Rev. Lett. {\bf 109}, 247604 (2012).


\bibitem{SP} A. A. Schafgans, K. W. Post, A. A. Taskin, Y. Ando, X.-Liang Qi, B. C. Chapler, D. N. Basov, Phys. Rev. B {\bf 85}, 195440 (2012).

 
\end{thebibliography}
\end{document}